# Trion Species-Resolved Quantum Beats in MoSe$_2$

Gabriella D. Shepard[1,2], Jenny V. Ardelean[3], Daniel A. Rhodes[3], X.-Y. Zhu[4], James C. Hone[3], and Stefan Strauf[1,2]

[1]Department of Physics, Stevens Institute of Technology, Hoboken, New Jersey 07030, United States

[2]Center for Distributed Quantum Computing, Stevens Institute of Technology, Hoboken, New Jersey 07030, United States

[3]Department of Mechanical Engineering, Columbia University, New York, New York 10027, United States

[4]Department of Chemistry, Columbia University, New York, New York 10027, United States

*Address correspondence to: strauf@stevens.edu

## ABSTRACT

Monolayer photonic materials offer a tremendous potential for on-chip optoelectronic devices. Their realization requires knowledge of optical coherence properties of excitons and trions that have so far been limited to nonlinear optical experiments carried out with strongly inhomogenously broadened material. Here we employ h-BN encapsulated and electrically gated MoSe$_2$ to reveal coherence properties of trion-species directly in the linear optical response. Autocorrelation measurements reveal long dephasing times up to T$_2$=1.16±0.05 ps for positively charged excitons. Gate dependent measurements provide evidence that the positively-charged trion forms via spatially localized hole states making this trion less prone to dephasing in the presence of elevated hole carrier concentrations. Quantum beat signatures demonstrate coherent coupling between excitons and trions that have a dephasing time up to 0.6 ps, a two-fold increase over those in previous reports. A key merit of the prolonged exciton/trion coherences is that they were achieved in a linear optical experiment, and thus are directly relevant to applications in nanolasers, coherent control, and on-chip quantum information processing requiring long photon coherence.

Monolayer transition metal dichalcogenides (TMDCs) are two-dimensional semiconductors that have gained considerable interest for optoelectronic[1–3] and valleytronic applications.[4–8] As a result of their low dimensionality and reduced dielectric screening, attractive Coulomb interactions are remarkably strong in these materials, resulting in large exciton ($X^0$) binding energies of up to several hundred meV.[9–11] While molybdenum-based TMDCs are direct bandgap semiconductors featuring bright excitons, it was recently shown that tungsten-based TMDC materials have dark excitonic states energetically lower than the optically bright exciton states at the K and K' points of the Brillouin zone, resulting in transitions to an indirect semiconductor at cryogenic temperatures.[12,13] Higher-order Coulomb correlations such as three-particle trion states that are either negatively charged ($X^-$) or positively charged ($X^+$) have also been observed with binding energies in the range of 20-36 meV.[11,14,15]

To realize many of the perceived applications, it is important to know how the strong Coulomb correlations and interactions of these monolayers with the environment affect the photophysics and coherence dynamics of excitons and trions. Since most studies have been carried out on TMDC material with inhomogenously broadened optical transitions with typical $X^0$ linewidths of 10-50 meV, nonlinear techniques based on four-wave mixing (FWM) had to be employed to gain insight into the underlying homogenous optical properties. These FWM experiments reveal relatively fast dephasing for both neutral excitons $T_2$= 200-620 fs in $MoS_2$, $WSe_2$, and $MoSe_2$[16–19] and trions (460-510 fs in $MoSe_2$[16,17]). With increasing temperature, electron-phonon interaction was found to be the limiting factor on exciton dephasing in all TMDCs that are either in monolayer or bulk form.[19] Recent calculations of the microscopic polarization indicate that in molybdenum-based TMDCs exciton coherence is limited by intravalley scattering with acoustic phonons while coherence in tungsten-based TMDCs suffers additionally from intervalley scattering into dark states.[12]

Among the molybdenum-based TMDCs that do not suffer from dark exciton states, $MoS_2$ was reported to display spectrally merged exciton and trion transitions even at 4 K,[20] while both transitions are clearly separated in $MoSe_2$. This makes $MoSe_2$ an ideal candidate to investigate coherent coupling among excitons via quantum beat spectroscopy,[21] similar to early studies of quantum wells and quantum dots that show coherent coupling between free and bound excitons[22] or light-hole and heavy-hole excitons.[23,24] More recently, it has been shown via FWM experiments that the exciton and trion in $MoSe_2$ can be coherently coupled but with rather fast decoherence time of about 250 fs.[16,17,25]

The question arises to what extent previous experiments on inhomogenously broadened optical transitions are able to determine intrinsic material properties, particularly since substrate interactions, native defects, or substitutional impurities might influence the dephasing properties. Early studies report for example spontaneous emission lifetimes $T_1$ ~200 fs in $WSe_2$ determined from analysis of FWM signals[18] while direct linear optical measurements with sufficient timing resolution reveal one order of magnitude longer values of $T_1$=2 ps for $X^0$ in $WSe_2$, as well as $T_1$=1.8 ps for $X^0$ and $T_1$=16 ps for the trion in $MoSe_2$.[26] Equally important is the layer transfer process that might leave polymer residue or incorporate nanobubble formation and thus exciton localization.[27] Recent experiments have demonstrated that it is possible to isolate from detrimental substrate effects and realize narrowband exciton emission linewidth near the homogenous limit directly in the linear optical emission by encapsulating the TMDC monolayer between hexagonal boron nitride (h-BN),[15,20,28] with the narrowest **X⁰** linewidth of 1.7 meV in our recent work.[28]

Reaching the homogenous linewidth in linear optical emission from TMDC materials now opens the door to studying decoherence properties directly in the time domain by recording the first-order autocorrelation function $g^{(1)}(\tau)$ with a Michelson interferometer, as was previously demonstrated for 0D quantum dots[29–32] or 1D carbon nanotubes,[33] but has not yet been applied to 2D materials. Coherent autocorrelation spectroscopy was also shown to be able to distinguish between Gaussian and Lorentzian emission profiles[29] and furthermore to be insensitive to spectral fluctuations (e.g. spectral diffusion) on time scales that are longer than the coherence time of the emitted light.[34]

Here we show that h-BN encapsulation of $MoSe_2$ leads to a transition from Gaussian to Lorentzian spectral lineshapes of the exciton emission in both time-integrated PL emission and interferometric dephasing time measurements. The narrow linewidth combined with electrical gating allows us to resolve the fine structure of spectrally distinct trion species **X⁺** and **X⁻** that are separated by 2-3 meV, depending on carrier density. We also apply linear optical autocorrelation measurements and find significantly prolonged dephasing times up to $T_2=1.16\pm0.05$ ps for **X⁺** due to the minimization of substrate-induced emitter dephasing. We also provide evidence that the **X⁺** forms from spatially localized hole states making this trion less prone to dephasing in presence of elevated hole carrier concentrations. Finally, we present trion species resolved quantum beat signatures directly in linear optical response revealing coherent coupling between excitons and trions that have a two-fold prolonged dephasing time up to $\tau_{XT} = 0.6$ ps compared to previous reports.

## RESULTS AND DISCUSSION

We use three types of samples to systematically investigate the influence of dielectric environment and carrier doping on the optical coherence properties of excitons in $MoSe_2$, including monolayer $MoSe_2$ directly on $SiO_2$ (BARE), monolayer $MoSe_2$ encapsulated between top and bottom layers of h-BN (STACK), and monolayer $MoSe_2$ encapsulated between layers of h-BN with additional top and back-gate graphene electrodes to realize voltage-controlled gating (GATE), as schematically shown in **Figure 1**. These samples were created through dry stamping techniques as described in the methods section. Under nonresonant laser excitation at 1.9 eV (650 nm) the PL spectrum from the BARE sample shows two well-separated but spectrally broad (10 meV) emission peaks corresponding to the $\mathbf{X^0}$ and $\mathbf{X^-}$ emission at 1.66 eV and 1.63 eV respectively, as shown in **Figure 1a**. Similarly, the PL spectrum for the STACK sample contains emission from $\mathbf{X^0}$ and $\mathbf{X^-}$ (**Figure 1b**) but at a 4-fold reduced full-width at half maximum (FWHM) linewidth value of 2.4 meV for $\mathbf{X^0}$, as well as with a lineshape that changed from Gaussian for the BARE sample to Lorentzian for the STACK sample. In addition, the PL emission energy of all transitions in h-BN encapsulated samples is shifted about 20 meV to lower energies, likely due to a combination of reduced strain[35–37] affecting the TMDC bandgap, as well as a change in dielectric constant of h-BN as compared to $SiO_2$, which effectively reduces the exciton binding energy. The spectrum of the GATE sample in **Figure 1c** was recorded under 2.33 eV (532 nm) laser excitation and a back-gate voltage ($V_{BG}$) set to 0 V, resulting in $\mathbf{X^0}$ and $\mathbf{X^-}$ transitions at 1.64 eV and 1.61 eV and FWHM values of 2.1 meV and 2.9 meV, respectively. Energy transitions and linewidth are quite comparable between the STACK and GATE sample with only minor variations, indicating that the introduction of the additional graphene electrodes does not alter the photophysical properties.

Electrical gating modulates the doping level of the TMDC material and is furthermore an ideal way to manipulate the contributions of neutral and charged excitons in the optical emission spectrum.[11] In this way, the GATE sample eliminates the need for additional spectral filtering in interferometric coherence dynamics studies. **Figure 2a** shows a color map of the emitted PL under fixed optical pump power as a function of $V_{BG}$ that varies from -14 V to +8 V. The three distinct optical transitions that dominate the spectrum in different gate voltage regimes with extracted peak intensities is displayed in **Figure 2b**. While the pump laser creates a fixed density of $\mathbf{X^0}$ the increasing positive gate voltage causes the $\mathbf{X^0}$ population to decrease since there are additional free electrons in the system (n-type) to form the $\mathbf{X^-}$ state with higher likelihood, and consequently only $\mathbf{X^-}$ emission is detectable at the highest voltage setting (+8V) and $\mathbf{X^0}$ is fully quenched (**Figure 3a**). Similarly, when a negative voltage is applied the material becomes *p*-doped due to excess holes, allowing for the formation of $\mathbf{X^+}$ while $\mathbf{X^0}$ is largely quenched (**Figure 3b**). Furthermore, gating the material in this direction results in a gradual increase of the $\mathbf{X^+}$ intensity by one order of magnitude despite the laser creating a fixed number of excitons. This behavior was observed to be fully reversible and accompanied by a minor photocurrent up to 0.5 nA at -14 V. When the laser is blocked, no electroluminescence was observed, excluding ambipolar carrier injection as a cause. This reversible PL enhancement is similar to our recent findings in ionic-liquid gated $MoSe_2$ and attributed to the effective screening and passivation of nonradiative recombination centers affecting the exciton intensity.[38] The increase in carrier concentration upon gating is also accompanied by distinct signatures in the spectral linewidth, as shown in **Figure 2c**. While the **X$^+$** linewidth is largely unaffected and varies by about 10% (1.8-2.0 meV) when gated from -1.9 V to -14 V, the **X$^-$** state broadens initially strongly from 2.4 meV to 3.6 meV, i.e. increases 50% when gated from -0.6 V to +1.1 V, followed by a slight decrease out to $V_{BG}$= +8 V that occurs simultaneous with the intensity bleaching of the **X$^0$** population. Likewise, the optical transition energies are largely unaffected in the interval from -11 V to +1 V, while stronger gating results in distinct redshifts for the trions of about 2 meV while **X$^0$** undergoes a blueshift of comparable magnitude (**Figure 2d**).

Importantly, one can clearly distinguish between the two trion species due to the narrow homogeneous linewidths observed in this sample, as is visualized in **Figure 2a.** The fine stepping sweep of $V_{BG}$ in **Figure 3c** reveals that **X$^-$** and **X$^+$** have distinctly different energies by 2-3 meV due to the underlying Coulomb interaction, which has previously been unobserved in this material in the presence of dominant inhomogeneous broadening.[11] Specifically, in the cross-over region, the separation varies from 2.8 meV to 3.1 meV (-1.1 to -0.9V) while near the endpoints when either **X$^-$** or **X$^+$** is gated out the value approaches 2.0 meV. The center of the cross-over from **X$^-$** to **X$^+$** occurs at $V_{BG}$ = -0.9 V where both states contribute equally to the emission spectrum. This gate voltage setting corresponds to the charge-neutrality point from an optical point of view, similar to the minimum conductance dip of the Dirac point in graphene field-effect transistors, that can also be observed away from $V_{BG}$=0 V in the presence of residual charge defects in the monolayer.[39] Therefore, at $V_{BG}$=0 the GATE sample is in the n-type doping regime with a dominant **X$^-$** that matches the observation that the BARE and STACK sample are dominated by the **X$^-$** state, i.e. the exfoliated $MoSe_2$ material is intrinsically n-type.[40]

The observed pronounced changes from an inhomogeneously broadened to a homogenous lineshape at significantly narrower FWHM values in h-BN encapsulated samples is expected to lead to longer exciton and trion dephasing times. To directly access the decay of coherence in the time domain ($T_2$) for the various exciton states we have recorded the first-order autocorrelation function g$^{(1)}(\tau)$ with a Michelson interferometer. The free-space Michelson interferometer depicted in **Figure 4a** inset consists of a 50/50 beamsplitter, a fixed mirror in one arm, and a second moveable mirror in the other to create a variable time delay. **Figure 4a** shows an exemplary interferogram that can be described by $g^{(1)}(\tau) \sim I(\tau) = I_0(1 + V(\tau)\cos[\omega t])$, where $I_0$ is the average counts per second, $V(\tau)$ is the visibility contrast, $\omega = E_0/\hbar$ the transition frequency, and $E_0$ the transition energy. A zoomed-in view of the region highlighted in orange demonstrates the distinct fringes that develop as the time delay is increased. The visibility contrast $V(\tau) = (I_{max} - I_{\min})/(I_{max} + I_{\min})$, i.e. the envelope function of the interferogram, is calculated here for groups of 10 fringes, where $I_{min}$ and $I_{max}$ are the fringe minima and maxima for each grouping. The $T_2$ time can be extracted from fitting the contrast decay as a function of the delay time $\tau$. To distinguish between the two distinct decay types, the visibility contrast from the interferogram is plotted on a semi-log plot, where a Lorentzian decay appears linear and a Gaussian appears quadratic, i.e. is characterized by a significant bowing in the initial part of the decay.

We first investigate the decoherence properties of $X^0$ and $X^-$ separately by only allowing light from one state into the interferometer. For the BARE and STACK sample the emission is spectrally filtered with a 10-nm bandpass before passing through the interferometer, and care is taken to ensure that the filter is centered over the emission peak to avoid artificial effects on the contrast decay. **Figure 4b** shows the visibility decay resulting from $X^-$ in the BARE sample. The pronounced bowing in the first 200 fs cannot be described by a monoexponential fit for this inhomogeneously broadened system. Instead, a Gaussian function $g^{(1)}(\tau) \sim \exp(-\tau^2/\tau_c^2)$ with a decay constant $\tau_c = 0.36 \pm 0.03$ ps best describes the observed decay dynamics of the $X^-$ state, which corresponds to $T_2 = \sqrt{\pi/2}\, \tau_c = 0.45 \pm 0.04$ ps. In contrast, performing this same measurement on the STACK sample produces strikingly different results as a result of much improved homogeneity, leading to a fringe decay that fits to a pure monoexponential decay function $g^{(1)}(\tau) \sim \exp(-|\tau|/\tau_c)$ with $\tau_c = T_2 = 0.73 \pm 0.04$ ps (**Figure S1**). Likewise, the fringe decay of $X^-$ for the GATE sample fits well to a monoexponential function with $T_2 = 0.71 \pm 0.04$ ps (**Figure 4b**). For the case of the GATE sample we did not utilize spectral filtering since $X^0$ can be fully gated out at $V_{BG} = +8$ V. The dephasing times of the $X^-$ state in the STACK and GATE sample match within the error, indicating that neither the spectral filtering, nor the application of a large positive gate voltage change the dephasing mechanism of $X^-$ considerably.

Interestingly, the $X^+$ state that is accessible in the GATE sample has a monoexponential decay that corresponds to a significantly longer dephasing time of $T_2 = 1.16 \pm 0.05$ ps, as shown in **Figure 4c**. Apparently, dephasing of positively charged excitons is governed by a microscopic mechanism that leads to prolonged coherence properties compared to both the $X^-$ state and the $X^0$ state that dephases with $T_2 = 0.98 \pm 0.05$ ps (see **Figure S1**). Assuming $X^+$ move freely within the 2D material they are expected to scatter at rates that are much faster than $X^0$ and thus have a reduced $T_2$ time and thus broader linewidth. The time-resolved measurements show that the contrary is true. Furthermore, **Figure 2c** indicates that the $X^0$ linewidth is always significantly broader than the $X^+$ linewidth, with the exception of the charge neutrality region. At highest hole concentrations, this difference in linewidth reaches a maximum with $X^0$ being 1.15 meV broader than $X^+$. This finding is similar to previous studies of positively charged excitons in modulation-doped semiconductor quantum wells and was attributed to a localization of the holes in potential fluctuations.[41] In the case of hole localization, the $X^+$ are immobile and less prone to scattering, which is also in agreement with the observed linewidth values that are largely unchanged for $X^+$ upon increasing hole concentration but broaden strongly for $X^-$ under increasing electron concentration (**Figure 2c**). That holes forming $X^+$ are localized is plausible given that the material is intrinsically n-type, possibly due to Se vacancies,[40] which leave behind positively charged immobile core ions ($D^+$) if the mobile electrons are gated out.

**Figure 5** compares the contribution of the trion dephasing time $T_2$ from the first-order autocorrelation experiments to the spectral linewidth in the BARE, STACK, and GATE configurations for both monolayer $WSe_2$ and $MoSe_2$. The most pronounced reduction of environment induced emitter dephasing is achieved by the h-BN encapsulation, which also changes the lineshape from Gaussian to Lorentzian. The smallest linewidth is achieved for the **X$^+$** state (1.2 meV) that is better protected from dephasing due to the hole localization. These values have been calculated including effects of spontaneous emission rate ($T_1$) and linewidth profiles. Specifically, the spectral linewidth $\Gamma$ is described by the well-known relation $\Gamma = \frac{2\hbar}{T_2}$ for the Lorentzian and $\Gamma = \frac{\hbar\sqrt{8\pi ln2}}{T_2}$ for the Gaussian case, where $\frac{1}{T_2} = \frac{1}{2T_1} + \frac{1}{T_2^*}$, and $T_1$ is the radiative lifetime of the emission and $T_2^*$ is the pure dephasing time.[29,42] While contributions from optical recombination to the spectral linewidth are rather small, given that $T_1$ values for trions in $MoSe_2$ reach up to 17 ps at 8 K,[26] i.e. $T_2 \sim T_2^*$, they are already included in the experimental $T_2$ values determined from the interferometer.

When comparing the linewidth contributions from $g^{(1)}(\tau)$ recorded from the interferometer to those determined from the PL lineshape analysis of time-integrated spectra, a clear energy difference $\Delta E$ becomes apparent between the two values (see **Figure 5**). While the $\Delta E$ values are largest for the BARE samples (7.6 meV) they reduce significantly in the GATE sample (0.8 meV). It is known that time-integrated lineshape measurements can be affected by dynamic disorder on ns-ms time scales, while the $g^{(1)}(\tau)$ at kHz count rates measures solely the self-interference of individual photons in the interferometer, i.e. are not sensitive to spectral fluctuations occurring on time scales that are longer than the coherence time of the light. As an example, we have recently shown that the $\Delta E$ value for excitons in carbon nanotubes can amount up to 5 meV if affected by substrate-induced ($SiO_2$) spectral diffusion, but can be virtually absent (<0.1 meV) in ultra-clean and air-suspended carbon nanotubes.[33] While the microscopic origin of dynamic disorder in TMDCs is not fully understood we note that telegraph noise from fluctuating charges in the exciton/trion vicinity gives rise to spectral broadening that can maintain a Lorentzian lineshape,[43] as we observe in GATE and STACK configurations. It was furthermore shown that excitons in $MoS_2$ undergo strong quantum-confined Stark shifts up to 16 meV when induced by an electric field perpendicular to the TMDC monolayer.[44] Likewise, small variations in the exciton or trion screening radius have been shown to significantly change their binding energies.[15] It is thus plausible that Stark-induced spectral diffusion and/or rapid charge fluctuation affecting the Coulomb binding energy can give rise to dynamic disorder broadening that is characterized by the measured $\Delta E$ values. As a result, the indirect method from the time-integrated PL spectrum cannot be used to get an accurate estimate on emitter dephasing times in 2D materials, but must be considered as a lower bound. Instead, the time-resolved experiments presented here determine values that are largely unaffected by dynamic disorder and thus relevant for cavity-QED experiments or nanolasers that are governed by ps time-scale coherences.[45]

We now investigate the decoherence properties for the case that light from both states **X$^0$** and **X$^+$** is sent simultaneously into the interferometer. The underlying exciton-trion coupling in $MoSe_2$ is known to create a four-level diamond system depicted in **Figure 6a,** including the coherently coupled state **XT**, that leads to the observation of quantum beats in the time-domain. Since this is an interference effect, measurements are taken at gate voltage settings of -1.25 V where **X⁰** and **X⁺** have equal intensities to maximize their interference contrast (**Figure 6b**). From the spectral separation that reflects the additional Coulomb energy in the system one can determine an **X⁺** binding energy of 24.3±0.5 meV at $V_{BG}$=-1.25 V. Note that the **X⁺** binding energy determined in this way varies with increasing hole concentration from 23.9 meV at $V_{BG}$=-0.8 V to 25.6 meV at $V_{BG}$=-12.5 V, due to the interplay of Pauli blocking and many-body interactions.[14] The resulting interferogram in **Figure 6c** shows a prominent beat pattern with 15 discernible fringes, which is significantly more than the 2-3 fringes that are typically observed in FWM studies.[16,17] To determine the **XT** state decoherence time ($\tau_{XT}$), the beat envelope is fit using the expression $g^{(1)}(\tau) = I_0(1 + c \exp(-|\tau|/\tau_{XT})$, resulting in a decay time of $\tau_{XT} = 0.60\pm0.05$ ps. Additionally, the quantum beat period ($B_{XT}$) for the **X⁰** and **X⁺** states can be determined either from the difference between fringe minima or maxima, as illustrated by the black dashed lines in **Figure 6e**. The mean value determined from the 15 fringe minima/maxima amounts to $B_{XT}$ =175±2 fs for **X⁰/X⁺**. This beat period corresponds to an energy separation of ΔE 23.6±0.7 meV that matches the value determined at $V_{BG}$=-1.25 V from the time-integrated PL spectrum.

Likewise, at $V_{BG}$=1V both **X⁰** and **X⁻** have equal intensity (**Figure 6b**) and give rise to a quantum beat spectrum that is completely out of phase with the **X⁰/X⁺** case after four oscillations (**Figures 6e**). The corresponding dephasing time of the **X⁰/X⁻** system amounts to $\tau_{XT} = 0.58\pm0.05$ ps (**Figure 6d**) with a beat period of $B_{XT} = 162\pm2$ fs, resulting in ΔE 25.5±0.7 meV that matches the value determined from the time-integrated PL spectrum (26.6±0.5 meV), indicating their coherent coupling.[16] With increasing free electron concentration, the **X⁻** binding energy determined from the spectral separation changes from 26.27 meV at $V_{BG}$=-0.9 V to 28.69 meV at $V_{BG}$=+5V. The **X⁻** state binding energy is thus significantly stronger affected than the **X⁺** state, i.e. shows a 2.4 meV variation over 6 V compared to 1.6 meV variation over 13 V, as one would expect if the **X⁺** are localized and thus less affected by many-body interactions. We further note that quantum beat oscillations in linear optics experiments are not unique to MoSe₂, but their onset can also be observed in monolayer WSe₂ in BARE configuration and at elevated temperatures of 80 K where the bound exciton emission is thermally quenched, albeit at significantly reduced values of only 100 fs for the exciton-trion coherence time (see **Figure S2**).

Previous work on monolayer MoSe₂ determined quantum beat dephasing times of $\tau_{XT}$=0.25 ps and exciton/trion dephasing times of $T_2$=0.47/0.51 ps, respectively, from nonlinear FWM in inhomogeneously broadened material.[17] In contrast, the coherence times determined in this work are about twice as long ($T_2$-$X^+$=1.1 ps, $\tau_{XT} = 0.60$ ps), as a result of h-BN encapsulation that effectively removes environment induced emitter dephasing. It was also demonstrated that beat signatures with up to 3 resolved fringes withstanding positive delays up to 0.7 ps,[16] while we show in **Figure 6c** resolved fringes withstanding out to 1.5 ps, which is likewise twice longer. Apparently, nonlinear FWM experiments that determine homogeneous dephasing values do not necessarily determine intrinsic decoherence properties in 2D materials, since the spatially extended exciton and trion wavefunctions give rise to pronounced substrate interactions, particularly if TMDCs are in contact with $SiO_2$. In contrast to FWM experiments, we determined here directly the coherences in the linear optical emission of the material, making them readily accessible for applications.

## CONCLUSIONS

**In summary**, we have shown that h-BN encapsulation of $MoSe_2$ leads to a transition from Gaussian to Lorentzian spectral lineshapes of the exciton emission in both time-integrated PL emission and interferometric dephasing time measurements. In addition, we have shown that positively and negatively charged excitons can be distinguished in electrically gated samples with an energy separation of about 2-3 meV, depending on carrier density, which has been previously unobserved in $MoSe_2$. Furthermore, the coherence times determined in this work have been significantly improved by the h-BN encapsulation that effectively removes substrate-induced emitter dephasing. The detailed analysis of gate-tunable intensity, linewidth, and energy position of excitonic transitions combined with dephasing studies provides evidence that the **$X^+$** state forms via spatially localized hole states making this trion less prone to dephasing in presence of elevated hole carrier concentrations, with dephasing times as long as $T_2=1.1$ ps. Trion species resolved quantum beat interferometry reveals coherent coupling between excitons and trions that have a two-fold prolonged dephasing time up to $\tau_{XT}=0.6$ ps compared to previous reports. A key merit of the demonstrated prolonged exciton/trion coherences in our work with h-BN-encapsulated material is that they were achieved in a linear optics experiment, i.e. they directly benefit applications of monolayer TMDCs in nanolasers, coherent control, and on-chip quantum information processing requiring prolonged coherence.

# METHODS

**SAMPLE FABRICATION: BARE:** $MoSe_2$ from HQ graphene was mechanically exfoliated onto a viscoelastic stamp (Gel-Film®) using Nitto Denko tape. Thin layers were identified by their optical contrast using an optical microscope and then stamped onto a standard Si wafer with 90 nm oxide where the surface was cleaned and made hydrophilic to improve adhesion between the flake and the substrate. To achieve this, the $SiO_2$ substrate was submerged in 30% KOH solution for 20 min and rinsed in DI water for 3 min. Stamping was done immediately after surface preparations were completed. Similar procedures were carried out for the bare $WSe_2$ sample (2D semiconductors). **STACK:** $MoSe_2$ monolayers are mechanically exfoliated from a commercially obtained bulk crystal (HQ Graphene) onto an $SiO_2$/Si substrate (with a 285 nm thick $SiO_2$ layer) using a combination of oxygen plasma treatment and heating. High-quality h-BN layers are mechanically exfoliated without heating and characterized with atomic force microscopy to identify flakes 20-30 nm thick with clean and ultra-flat areas 20 ~ 25 µm in size to minimize sample disorder. Encapsulated h-BN/$MoSe_2$/h-BN samples were prepared using a dry transfer method with a PC/PDMS lens (PC - poly(bisphenol A carbonate) & PDMS - poly(dimethyl siloxane)) for sequential pickup of the flakes. The pickup process was carried out slowly, with heating/cooling at a rate of ~ 0.5°C/min and a vertical translation rate of ~0.25 µm/s to minimize cracks and trapped bubbles at the h-BN/$MoSe_2$ interface. **GATE:** Sample fabrication follows the same procedure as the STACK sample, but with an additional pickup step for the graphene layer. The graphene layers are exfoliated from Kish graphite (NGS) using the same procedure outlined for the $MoSe_2$ monolayers. The gate electrodes were defined by electron beam lithography followed by Au evaporation and lift off.

**OPTICAL MEASUREMENTS:** Micro-photoluminescence ($\mu$-PL) were taken inside a closed-cycle He cryostat with a 3.8 K base (attodry1100). Samples were excited with a laser diode operating at 650 nm or 532 nm in continuous wave mode. An Abbe limited laser spot size of about 500 nm was achieved using a cryogenic microscope objective with numerical aperture of 0.82. The relative position between sample and laser spot was adjusted with cryogenic piezo-electric *xyz*-stepper (attocube). Regular PL spectra from the sample were collected in a single-mode fiber, dispersed using a 0.75 m focal length spectrometer with a 300-groove grating, and imaged by a liquid nitrogen cooled silicon CCD camera. Laser stray light was rejected using either a long-pass 700 nm filter or a long-pass 550 nm filter. Light sent into the Michelson interferometer was recorded with a single photon counting silicon avalanche photodiode (Si APD) collected after the output port beam splitter.

# ACKNOWLEDGMENT

We like to thank Tony F. Heinz and Yuping Huang for fruitful discussions. S.S. and J.H. acknowledge financial support by the National Science Foundation (NSF) under the collaborative awards DMR-1506711 and DMR-1507423, respectively. S.S. acknowledges financial support for the attodry1100 system under NSF award ECCS-MRI-1531237. J.H. and X.-Y. Z. acknowledge support by the NSF MRSEC program through the Columbia in the Center for Precision Assembly of Superstratic and Superatomic Solids (DMR-1420634).

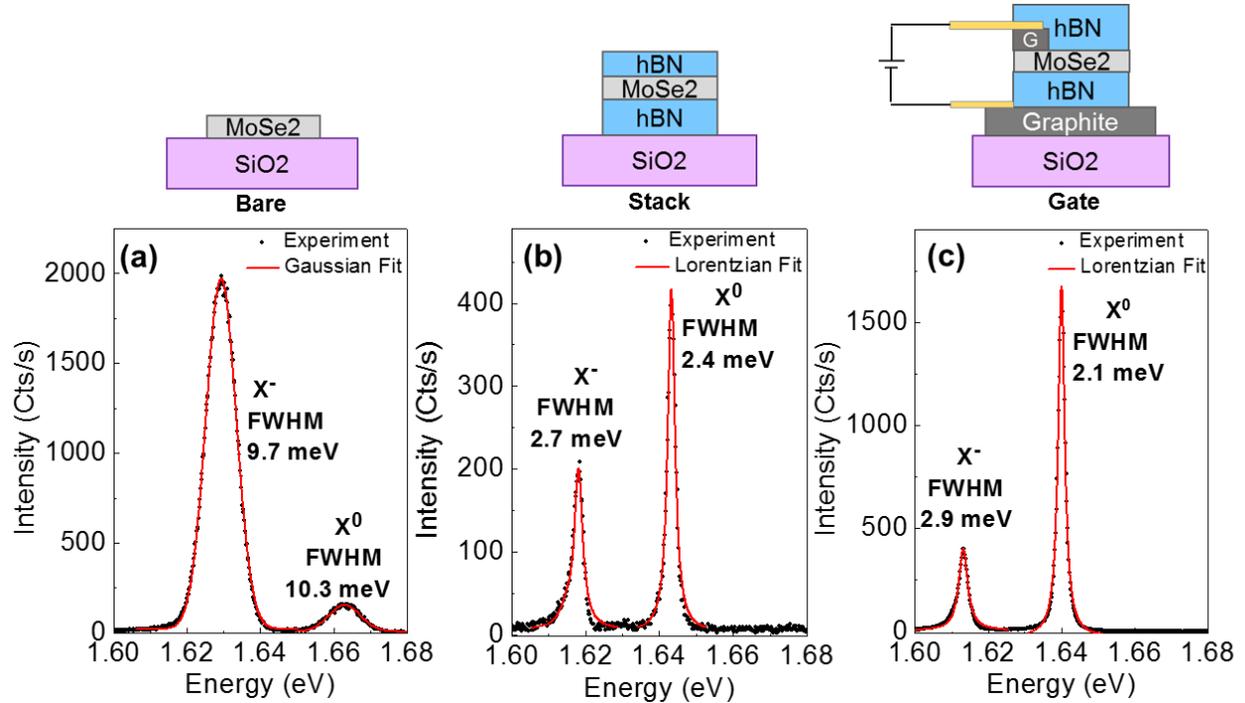

**Figure 1. Optical emission from bare and h-BN-encapsulated MoSe$_2$ at 3.8K.** Photoluminescence (PL) spectra recorded under nonresonant excitation for (a) MoSe$_2$ directly in contact with a 90 nm SiO$_2$ substrate (BARE), (b) h-BN-encapsulated MoSe$_2$ (STACK), and (c) h-BN-encapsulated MoSe$_2$ with additional few layer graphene backgate electrode, as well as monolayer graphene top contact, covering only the edges of MoSe$_2$ (GATE). The backgate voltage was set to 0 V. Note that GATE measurements are taken on regions of the material away from the graphene electrodes, which are known to quench PL. Each spectrum shows spectrally well-separated neutral exciton (**X$^0$**) and negatively charged exciton (**X$^-$**) emission. The h-BN-encapsulation significantly reduces the spectral linewidth. While excitons in (a) fit best by an inhomogeneous Gaussian lineshape (red solid line) the h-BN encapsulated cases (b,c) fit well to single Lorentzians. All data recorded at 3.8 K.

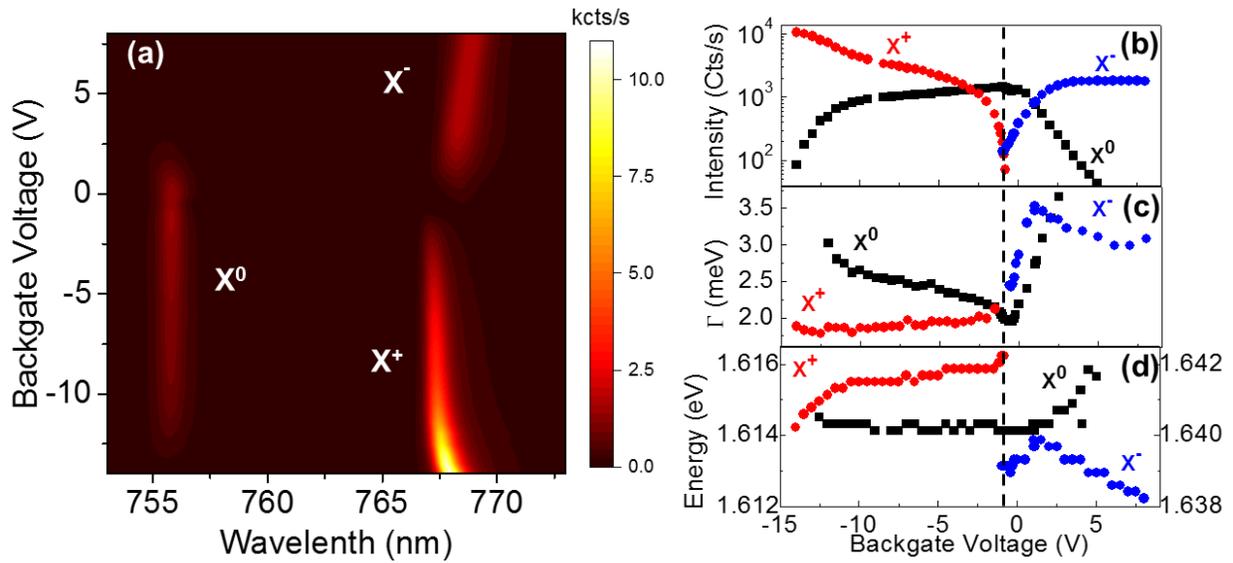

**Figure 2. Gate-tunable trion species in MoSe$_2$.** (a) MoSe$_2$ PL is plotted as a function of backgate voltage to demonstrate electrostatic control of exciton species. Two distinct trion species **X$^+$** and **X$^-$** are visible in addition to the neutral exciton **X$^0$**. (b) PL peak intensity at fixed laser pump power as a function of V$_{BG}$ with a step size of 0.5 V. Vertical dashed line highlights the charge-neutrality point at V$_{BG}$=-0.9 V where both trion species coexist at equal intensity. (c) Backgate sweep of the FWHM linewidth value of the exciton and distinct trion species. (d) Backgate sweep of the exciton and trion transition energies. Note that **X$^0$** energy values are plotted on the right y-axis and **X$^+$** and **X$^-$** values are plotted on the left y-axis. All data recorded at 3.8 K.

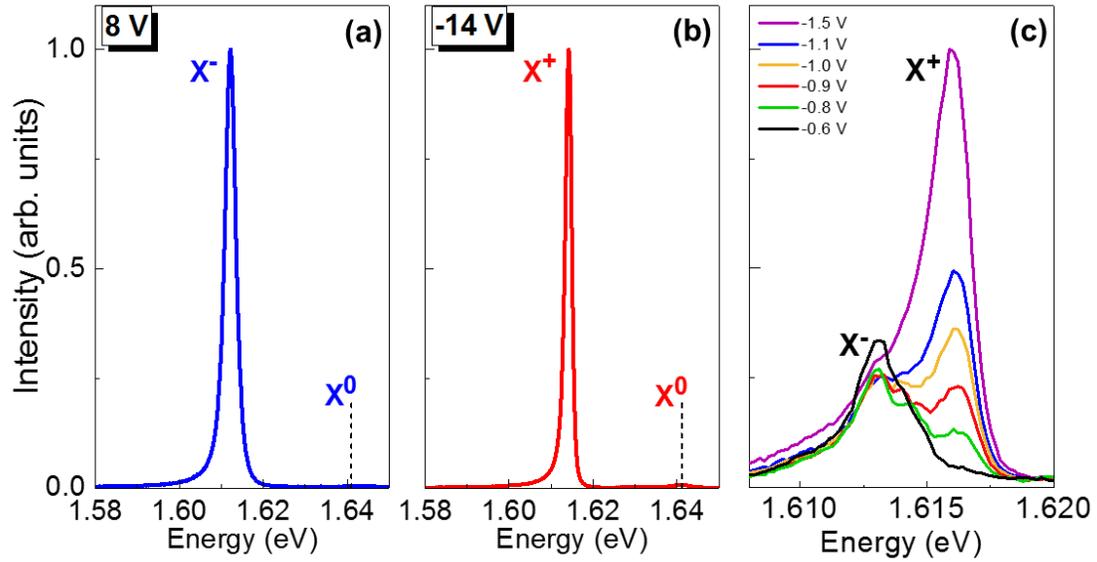

**Figure 3. Select PL spectra from backgate sweep** (a) Normalized PL spectra for $V_G$=+8V where the system is n-doped and $X^-$ dominates as a singular peak. The $X^0$ emission is fully quenched. (b) Normalized PL spectrum at $V_G$=-14V where the system is p-doped. The $X^+$ dominates as a singular peak while $X^0$ emission is largely quenched. Both spectra are recorded under nonresonant excitation (532 nm) with the same excitation power and have subtle asymmetric tales on their low energy sides deviating only slightly from a Lorentzian lineshape. (c) PL spectra fine-scan in steps of 0.1 V that captures the cross-over between the separate trion species $X^+$ and $X^-$.

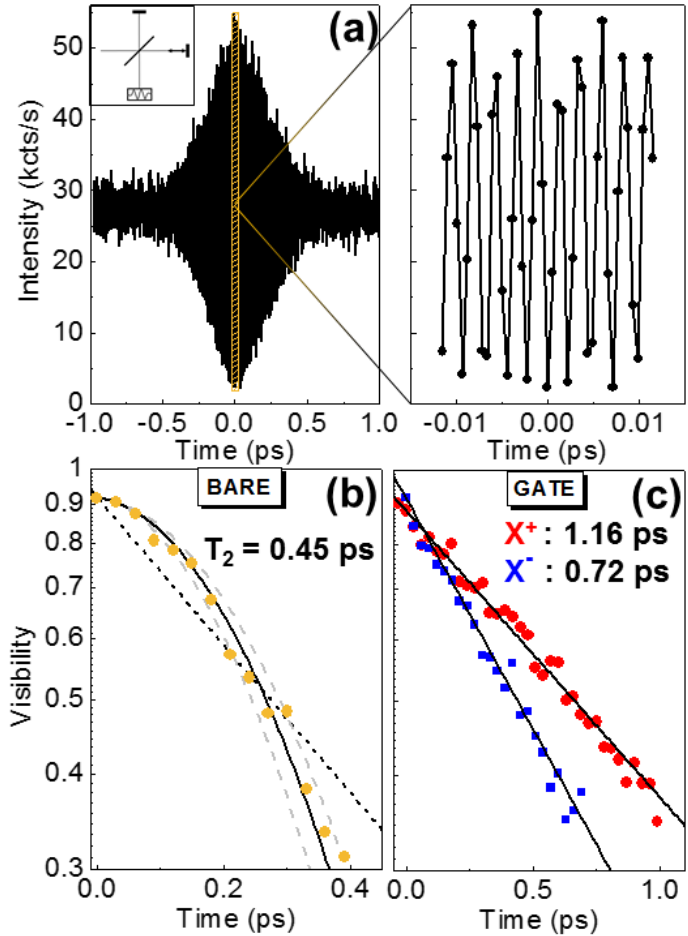

**Figure 4. Trion dephasing time measurements using a balanced Michelson interferometer.** (a) Interferogram of the first order coherence function $g^{(1)}(\tau)$. A zoomed-in version of the region highlighted in orange illustrates that fringes with near-unity contrast develop. Minimum and maximum values of 10 fringes are used to calculate the fringe visibility values. (b) The fringe visibility for the **X⁻** emission from the BARE sample (yellow circles) strongly deviates from a mono-exponential decay (black dashed line) revealing an inhomogeneous decay that follows a Gaussian fit (solid black line). Dashed gray lines indicate the spread on the fitted **t_c** value and the dashed black line illustrates monoexponential decay for comparison. (c) In contrast, the visibility for the **X⁻** (blue squares) and **X⁺** (red circles) emission from the GATE sample decay purely mono-exponential (solid black lines). The **T₂** dephasing time is determined from the decay constant **t_c** of each fit function. All data recorded at 3.8 K.

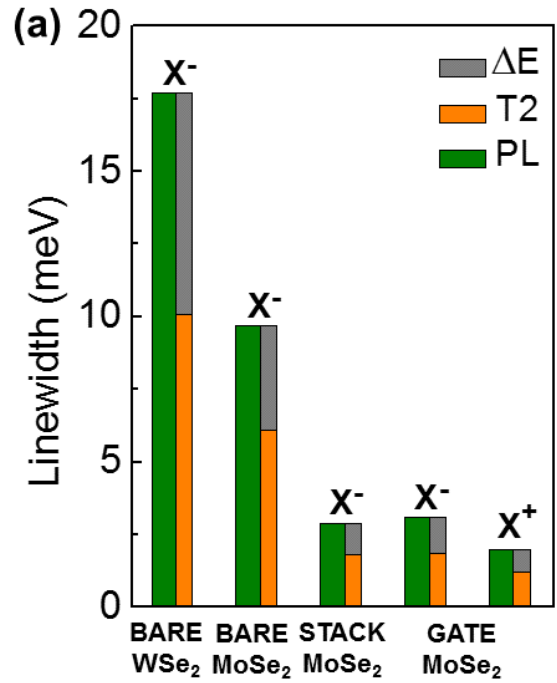

**Figure 5. Comparison of X⁻ linewidth from time-integrated and time-resolved measurements.** Energy difference (ΔE) between the X⁻ linewidth determined from time-integrated PL spectra (green) and the corresponding linewidth value determined from the time-resolved first-order autocorrelation function recorded with the Michelson interferometer (orange). The last column shows in addition the comparison for the X⁺ state for the GATE sample that has the narrowest linewidth and longest $T_2$ time. ΔE values are 7.6, 3.6, 1.1, 1.25, and 0.8 meV, respectively. All data recorded at 3.8 K.

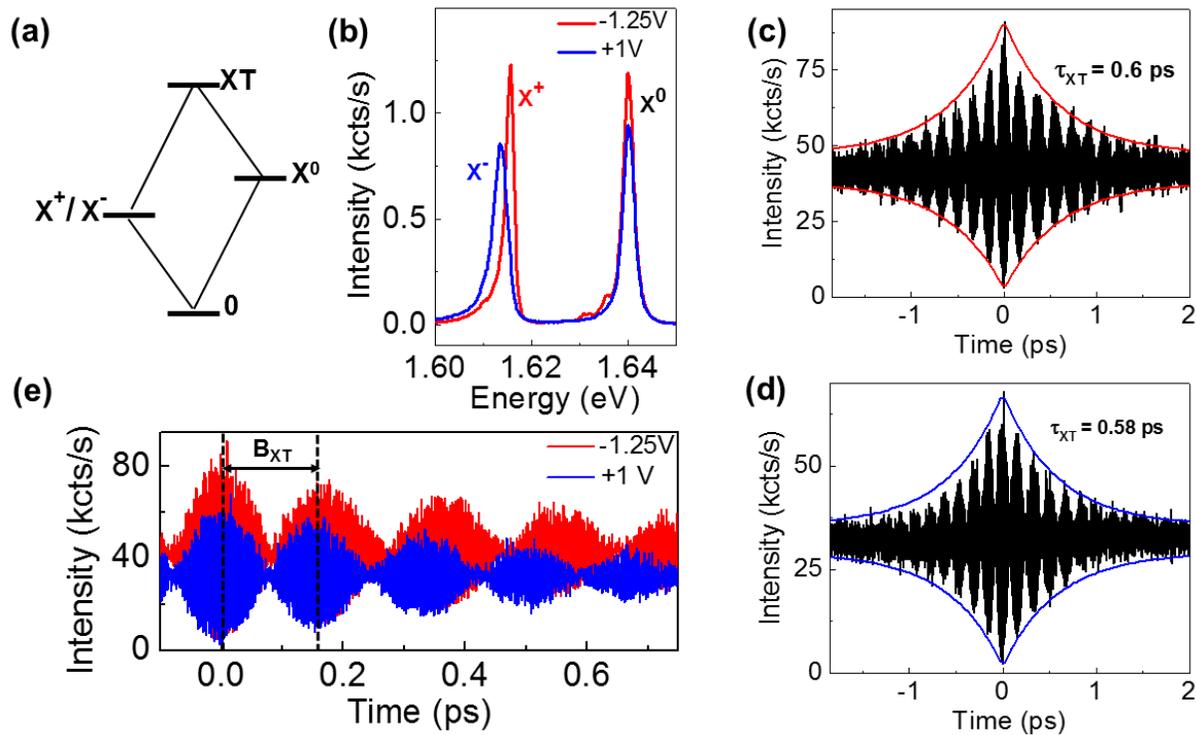

**Figure 6. Trion species resolved quantum beating in MoSe$_2$** (a) Four-level diamond system illustrating the coherently coupled exciton-trion state. (b) PL spectra recorded at gate voltage settings that produce equal intensities between **X$^0$** and **X$^+$** (-1.25 V) as well as **X$^0$** and **X$^-$** (+1 V) to maximize fringe contrast. (c) Interferogram produced by sending both **X$^+$** and **X$^0$** emission into the interferometer. Red line is a fit to the decay of the beat envelope corresponding to $\tau_{XT}$=0.60±0.05 ps for the exciton-trion coherence. (d) Same for the **X$^0$/X$^-$** case resulting in $\tau_{XT}$=0.58±0.05 ps. (d) Quantum beat spectra comparing **X$^+$** and **X$^-$** species that are initially in phase at time zero and develop out-of-phase after four beat periods. The quantum beat period B$_{XT}$ can be extracted from either the constructive interference maxima or the destructive interference minima, as indicated by the black dashed lines. All data recorded at 3.8 K.

# Trion Species-Resolved Quantum Beats in MoSe$_2$

## Supporting Information

Gabriella D. Shepard[1,2], Jenny V. Ardelean[3], Daniel A. Rhodes[3], X.-Y. Zhu[4], James C. Hone[3], and Stefan Strauf[1,2]

[1]Department of Physics, Stevens Institute of Technology, Hoboken, New Jersey 07030, United States

[2]Center for Distributed Quantum Computing, Stevens Institute of Technology, Hoboken, New Jersey 07030, United States

[3]Department of Mechanical Engineering, Columbia University, New York, New York 10027, United States

[4]Department of Chemistry, Columbia University, New York, New York 10027, United States

*Address correspondence to: strauf@stevens.edu

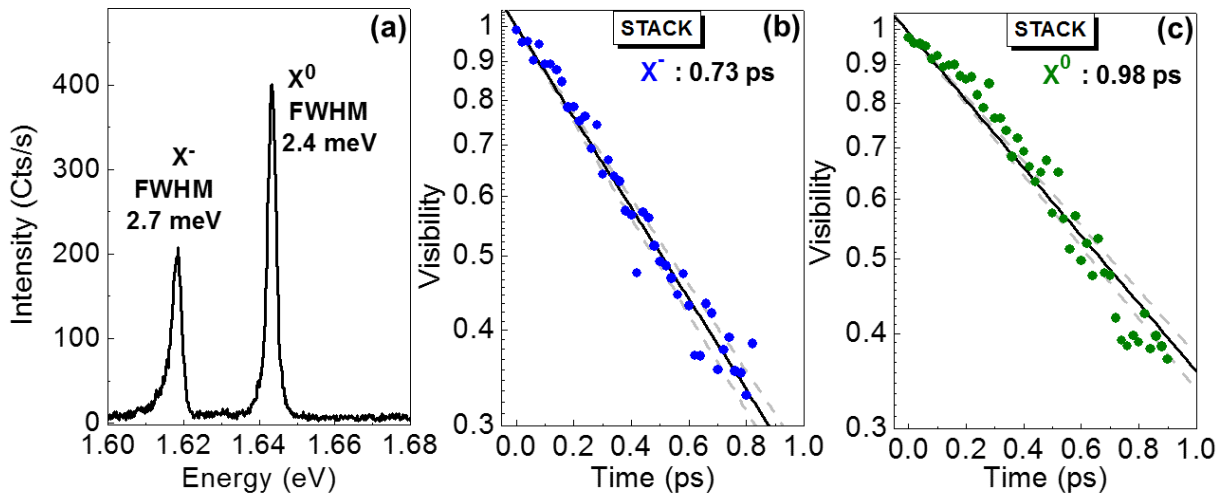

**Figure S1.** STACK dephasing time (T$_2$) measurements. (a) The PL spectrum for **X$^0$** and **X$^-$** recorded under nonresonant excitation (650 nm). Each exciton species is filtered through a 10-nm bandpass filter to study the individual T$_2$ times before being sent to the interferometer. The resulting fringe visibility for (b) the **X$^-$** and (c) **X$^0$** emission. Both follow a mono-exponential (solid black lines) decay. All data recorded at 3.8 K under the same excitation power.

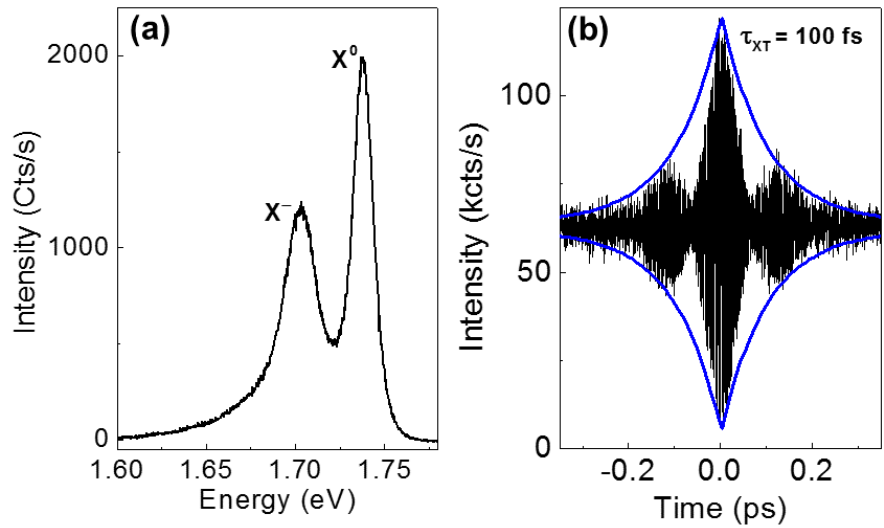

**Figure S2**. Quantum beats in WSe$_2$. (a) 80K PL spectrum for WSe$_2$ directly in contact with SiO$_2$ substrate recorded under nonresonant excitation (532 nm). Neutral exciton (**X$^0$**) and negatively charged exciton (**X$^-$**) emission show small spectral overlap in this material. (b) Resulting interferogram for sending **X$^0$** and **X$^-$** simultaneously through the interferometer. The blue line is a fit to the decay of the beat envelope corresponding to $\tau_{XT}$=0.10±0.05 ps for the exciton-trion coherence.